\documentclass[%
 reprint,
superscriptaddress,
showpacs,
nofootinbib,
 amsmath,amssymb,aps,
prd,
floatfix,
]{revtex4-1}

\usepackage{lipsum} 
\usepackage{blindtext}

\newcommand{\cmmnt}[1]{}

\usepackage{bm}
\usepackage{comment}
\usepackage{floatrow} 
\pdfoutput=1
\usepackage{amsmath,amsfonts,amsthm}
\usepackage{esdiff}  
\usepackage{booktabs}  
\usepackage{url}  
\usepackage{hyperref}  
\usepackage{relsize}
\usepackage{cleveref}  
	\crefname{equation}{equation}{equations}
	\crefname{figure}{figure}{figures}	
	\crefname{table}{table}{tables}
\usepackage[caption=false]{subfig}

\usepackage[normalem]{ulem}

\usepackage[usenames,dvipsnames,svgnames,table]{xcolor}

\usepackage[export]{adjustbox}
\usepackage{graphicx}

\usepackage{aurical}
\usepackage[T1]{fontenc}

\usepackage[sc]{mathpazo} 
\usepackage[T1]{fontenc} 
\usepackage{microtype} 

\usepackage{booktabs} 
\usepackage{float} 
\usepackage{hyperref} 

\usepackage{lettrine} 
\usepackage{paralist} 

\usepackage{titlesec} 
\renewcommand\thesection{\Roman{section}} 
\renewcommand\thesubsection{\Alph{subsection}} 
\titleformat{\section}[block]{\large\scshape\centering\bfseries}{\thesection.}{1em}{} 

\titleformat{\subsection}[block]{\scshape\centering}{\thesubsection.}{1em}{} 

\DeclareCaptionFormat{myformat}{#1#2#3\hrulefill}
\usepackage{float}
\floatstyle{plaintop}
\restylefloat{table}

\begin{document}
\nocite{TitlesOn}

\title{The Three Little Pigs and the Big Bad Wolf \\ \textit{Case studies of peer review}} 

\author{\href{http://www.amazon.com/author/evearmstrong}{Eve Armstrong}}

\email{evearmstrong.physics@gmail.com}

\affiliation{Department of Physics, New York Institute of Technology, New York, NY 10023, USA}
\affiliation{Department of Astrophysics, American Museum of Natural History, New York, NY 10024, USA}
\affiliation{\url{http://www.amazon.com/author/evearmstrong}}

\date{\today}
\date{April 1, 2022}

\begin{abstract}
I present for your appraisal three independent cases of the manuscript referee process conducted by a venerable peer-reviewed scientific journal.  Each case involves a little pig, who submitted for consideration a theoretical plan for a house to be constructed presently, in a faraway land.  An anonymous big bad wolf was assigned by the journal to assess the merit of these manuscripts.  The pigs proposed three distinct construction frameworks, which varied in physical and mathematical sophistication.  The first little pig submitted a model of straw, based on the numerical method of toe-counting.  His design included odd features, such as spilled millet and cloven-hoofprints on the window sill -- possibly a ploy to distract the wolf from the manuscript's facile mathematical foundation.  The second little pig used a more advanced approach, employing Newton's classical laws of motion, to propose a house of sticks.  This pig included in her manuscript copious citations by a specific wolf, possibly aiming to ensure acceptance by flattering the wolf whom she anticipated would be the referee.  The third little pig described an ostentatious house of bricks based on an elaborate dynamical systems and stability analysis, possibly scheming to dazzle and impress.  The big bad wolf did not appear moved by any of the pigs' tactics.  His recommendations were, for straw: the minor revision of water-proofing; for sticks: the major revision of fire-proofing, given concerns surrounding climate change; for bricks: unequivocal rejection, accompanied by multiple derogatory comments regarding "high-and-mighty theorists."  I describe each case in detail, and suggest that the wolf's reports might be driven as much by self interest as the manuscripts themselves -- namely, that at the time the wolf wrote the reviews, he was rather hungry.  Finally, I examine morals learned, if any.\\

\noindent
\textit{The characters portrayed in this fable are fictional.  Any resemblance to actual persons, living or dead, might be purely coincidental.}
\end{abstract}

\maketitle 

\section{Introduction} \label{sec:intro} 

\begin{figure}[H]  
  \includegraphics[width=0.7\textwidth]{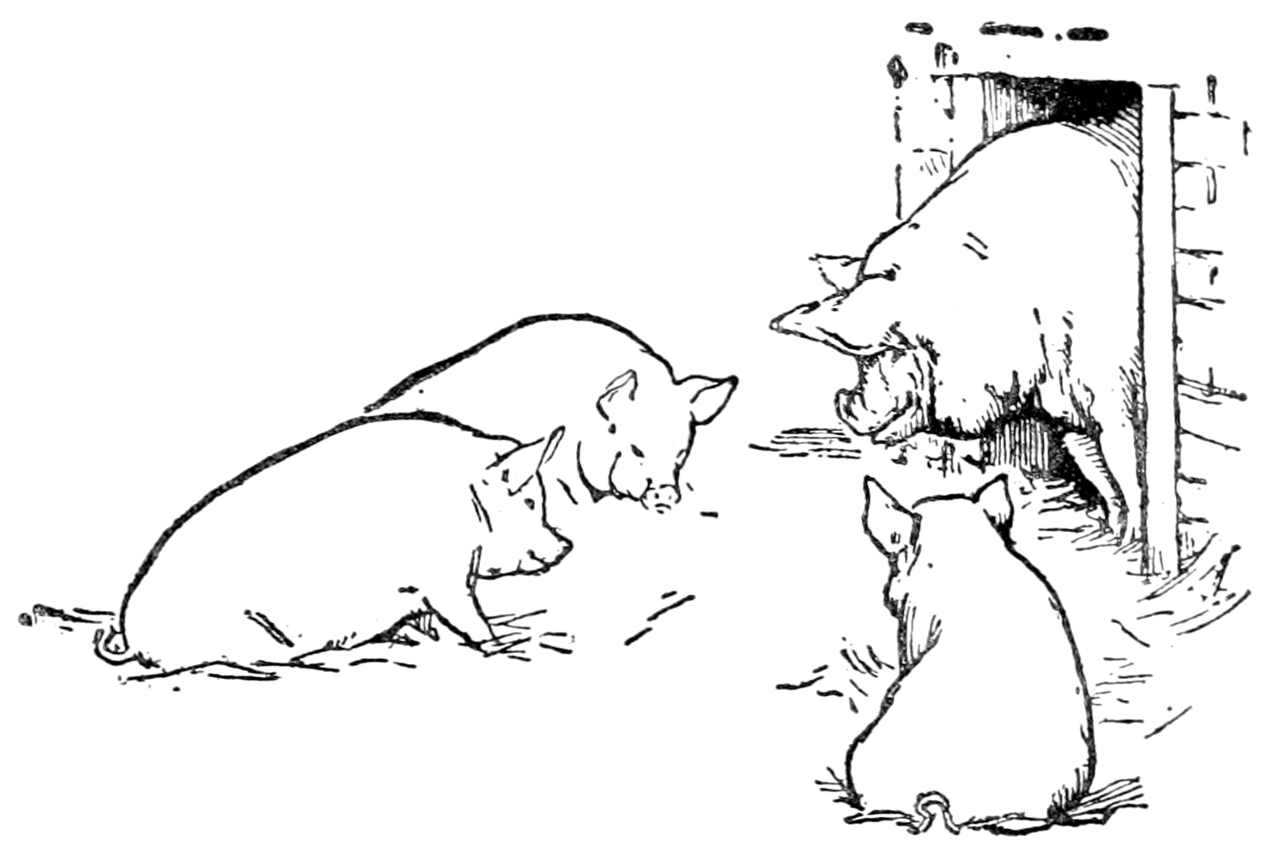}\\
  \caption{\textbf{The three little pigs with their mother, in happier times~\cite{pigs}.}}
\label{fig1}
\end{figure}

{\Fontamici{\Huge{O}\Large{nce upon a time in a faraway land}}}~\cite{hawking2010nature}, there lived an aging mother pig who had three little pigs (Fig.~\ref{fig1}), and she could no longer afford to care for them.  Thus she sent her children out into the world to forge their own livelihoods.  

\begin{figure}[H]  
  \includegraphics[width=\textwidth]{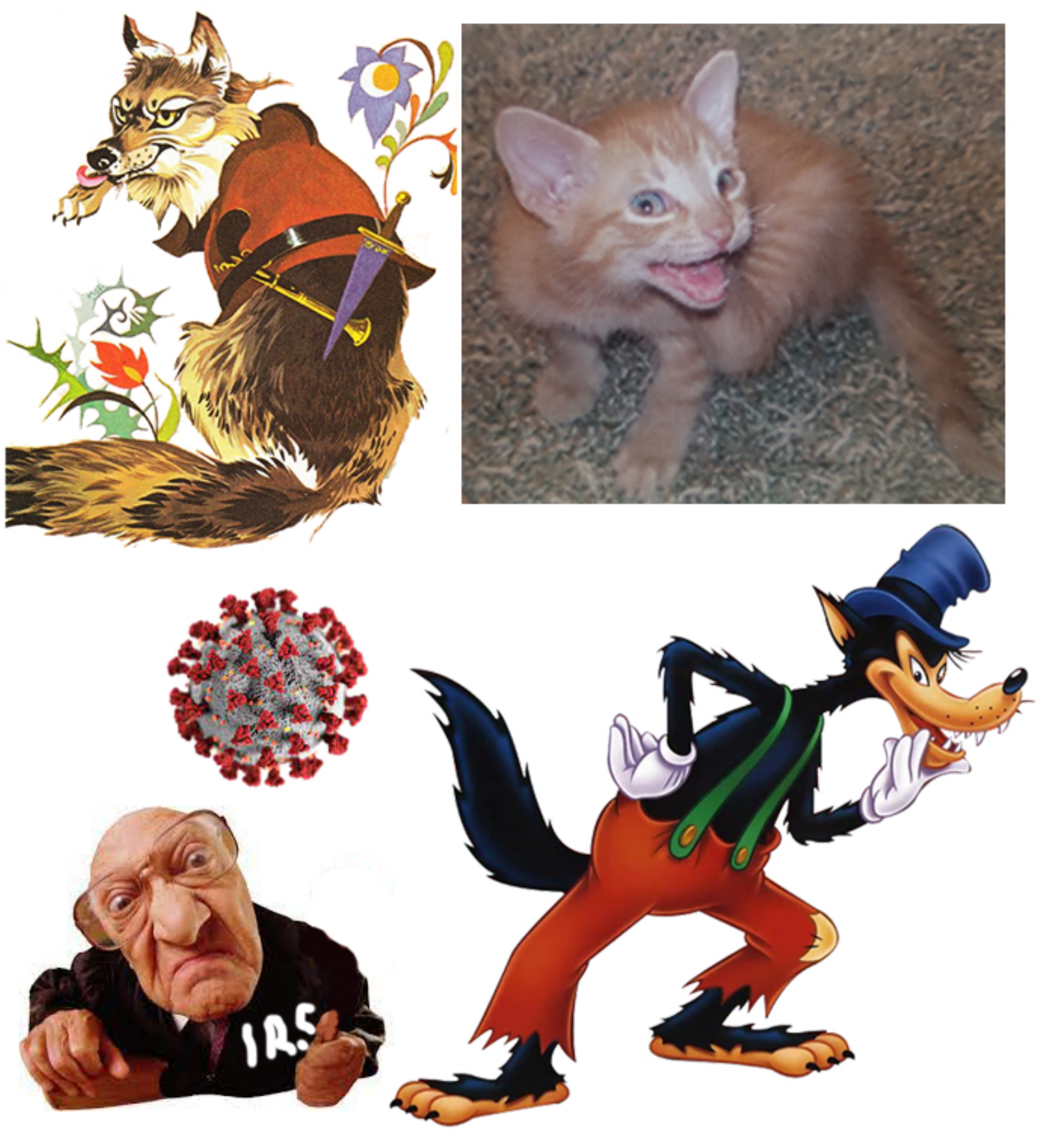}\\
  \caption{\textbf{Various visual interpretations of the anonymous big bad wolf, depending on whom one asks.}  \textit{Clockwise from Top Left}: Artistic Interpretation 1~\cite{wolf1}, kitten~\cite{sam}, Artistic Interpretation 2~\cite{wolf2}, IRS auditor~\cite{IRS}, novel coronavirus SARS-CoV-2~\cite{coronavirus}.}
\label{fig2}
\end{figure}

The pigs' mother had counseled them that in order to survive, one must construct a sound home, and broadcast its existence throughout the land as evidence of one's industriousness and intrinsic value.  To this end, each pig designed a home construction plan and submitted it for publication to a scientific journal.  For legal reasons, this article will not name the journal; let it suffice that it is a well known and respected non-open-access \cmmnt{ADD FOOTNOTE?  "To assess an open-access journal would open a can of worms that I do not have the energy to confront. See, for example, Footnote X.)"}peer-reviewed series.  The journal assigned an anonymous big bad wolf (Fig.~\ref{fig2}) to review the pigs' manuscripts.

The scholarly peer review process, or "refereeing," is academia's means of quality control in publication; it dates possibly to 1731~\cite{benos2007ups}.  The modern procedure consists of the assessment -- by referees with appropriate expertise and no competing interests -- of work to determine suitability for publication.  Suitability encompasses validity, quality, and originality.  The referee is usually anonymous, and advises the publication's editor on whether to accept the work, accept it following revisions, or reject it entirely.  At its best, peer review motivates authors to submit high-quality material, and then improves that material further, thereby maintaining the integrity of science.  At least, usually it filters out the truly bat-s\#\%t crazy stuff~\cite{bohannon2013s}.

In what follows, I present to you each little pig's manuscript in turn, followed by the wolf's comments.  Then we shall examine possible sources of bias inherent in the process.  Finally, we shall contemplate the moral of the story; that is, whether in the end, the characters have grown any the wiser.

\section{The First Little Pig} \label{sec:model}

\subsection{\textbf{Design and Method}}

\begin{figure}[H]  
  \includegraphics[width=0.3\textwidth]{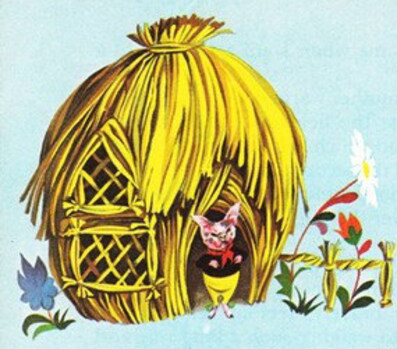}\\
  \caption{\textbf{First little pig's design for a house of straw}~\cite{houseStraw}.  Referee's main critique: "Not waterproof."  Recommended minor revision followed by acceptance.}
\label{fig3}
\end{figure}

The first little pig proposed a house of straw~\footnote{In fact, his original idea had been to apply Rumpelstiltskin's method of spinning straw into gold~\cite{grimm1812} -- arguably a more robust construction material.  Alas, he failed to crack Rumpelstiltskin's algorithm.} with a simple circular layout (Fig.~\ref{fig3}).  By most journal standards, it was a rudimentary plan.  Its sole mathematical content was an estimate of the number of bales of straw that would suffice to contain a comfortably-sized mud pit.   

Possessing a weak background in geometry, the pig found himself at a loss for calculating the circumference of a mud-pit-containing straw wall, based on the diameter of the mud pit.  To determine the required quantity, then, he opted to simply count to a very high number.  To perform this counting, the pig employed his toes.  

Pigs have four toes on each of their front cloven hooves~\cite{smith2012dissection}, two of which (the front two) a pig is able to see.  Further, as pigs have excellent memory~\cite{kouwenberg2009episodic}, this little pig found that he could perform the count twice, bringing his total to eight~\footnote{Let us assume that the pig's "eight" ascribes to the base-ten number system credited to Indian mathematicians in the seventh century~\cite{smith1911hindu}, although the pig did not specify this in his manuscript.}  (Eq.~\ref{eq1}).  After two full counts the pig found his memory faltering, and noted frustration over failing to count to the "very high number" described in his method.  He suspected that most numbers greater than eight are probably very high; however, for good measure, after including the eight, he then piled on as many more bales as he felt able to before growing tired.  He then concluded that, by eye, his final pile appeared sufficient~\footnote{Pigs have excellent spatial cognitive abilities~\cite{held2009advances}.}.
\begin{equation} \label{eq1}
[1,2,3,4,5,6,7,8] \hspace{0.5cm} (\text{toe count}).
\end{equation}

Now, setting aside the straw base, this pig's model included some odd features.  He added into the design spilled millet and barley grain in the mud pit, and cloven-hoofprints covering the window sill.  These features did not enhance the plan; on the contrary, they seemed to be blemishes, albeit harmless ones.  This is puzzling; see \textit{Discussion.} 

\subsection{\textbf{Referee Report 1}}

The big bad wolf's report to the first little pig appeared cursory.  He\footnote{While the big bad wolf might be male, female, or non-gender-conforming, for simplicity hereafter I adopt the pronoun "he."} did not note the spilled grains or cloven-hoofprints.  Nor did he comment on the pig's claim that all numbers above eight are large, which suggests that this wolf is not much of a mathematician either.  Paradoxically, most of his report consisted of a glowing commendation of the pig's "aesthetically pleasing simplicity of design" -- a comment that embodies the very soul of a mathematician.
\setlength{\tabcolsep}{5pt}
\begin{table*}
\small
\centering
\begin{tabular}{c | c | c} \toprule
 \textit{Law} & \textit{Statement} & \textit{Second little pig's claimed application}\\\midrule \hline\hline
 I & $\bm{F}_{net} = 0 \leftrightarrow \frac{d\bm{v}}{dt} = 0$ & "The house will not fall down all by itself."\\
 \tiny{II} & \tiny{($\bm{F}_{net} = m \bm{a}$)} & \tiny{("Irrelevant.")}\\
 II & $\bm{F}_{AB} = -\bm{F}_{BA}$ & "If the wind blows, the house will blow right back." \\\bottomrule
\end{tabular}
\caption{\textbf{Second little pig's use of Newton's three laws of motion, to establish the stability of a sticks-based house.}}
\label{table1}
\end{table*}

The wolf did acknowledge a possible personal disqualification for judging the soundness of the straw building material: namely, that he has lived the majority of his life in San Diego, California, and thus has experienced little weather of any kind.  On rare occasions he has endured rain, which compelled him to recommend the minor revision that the straw be water-proofed~\footnote{The wolf also noted that the pig at times confused the usage of the words "that" and "which," which is evidently on the list of this wolf's gripes with scientific literature.}.  The wolf recommended acceptance of the manuscript following that revision.

\section{\textbf{The Second Little Pig}}

\subsection{\textbf{Design and Method}}

\begin{figure}[H]  
  \includegraphics[width=0.5\textwidth]{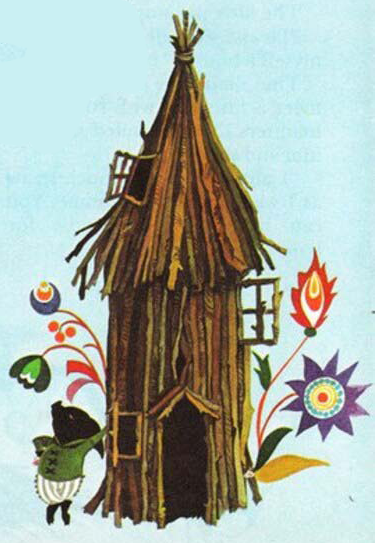}\\
  \caption{\textbf{Second little pig's design for a house of sticks}~\cite{houseSticks}.  Referee's main critique: "Not fireproof, which betrays ignorance of climate change."  Recommended major revision.}
\label{fig4}
\end{figure}

The second little pig proposed a house of sticks (Fig.~\ref{fig4}), adopting Sir Isaac Newton's three laws of motion for Earth-based dynamics~\cite{newton1934principia} to establish the spatio-temporal stability of the sticks (Table~\ref{table1}).  Newton's first law, commonly known as the "law of inertia," states that an object at rest will remain at rest unless acted upon by an external force.  The pig invoked this law to state that the house would not fall down of its own ambition.  

Newton's second law states that \textit{if} an external force were to act upon the house, then the house would undergo a nonzero acceleration.  The pig acknowledged this law, albeit in a font half the size as that of the rest of the manuscript, and made no comment on it (see \textit{Discussion}).

Newton's third law states that if two objects $A$ and $B$ touch each other, then Object $A$ exerts a force $F_{AB}$ upon Object $B$ that is equal and opposite to the force $F_{BA}$ that Object $B$ exerts on $A$.  The pig wielded this law to assert that if the wind were to blow upon the house of sticks, that the house of sticks would blow right back and thereby remain stationary. 


Now, akin to the first pig's spilled grain and hoofprints, this manuscript also sported an oddity: this time in the list of references.  The references included multiple citations, \textit{ad nauseum}, of papers by one particular wolf.  This pig's neglect of Newton's Second Law having heightened my senses, I investigated this matter.  A quick literature review revealed that the author this pig cited so copiously happens to be the wolf who is perpetually fraternizing in the woods trying to seduce Little Red Ridinghood~\cite{perrault1697}.  That wolf is well regarded for his research on home building material, possessing exhaustive experience with breaking into grandmothers' cottages.  This little pig's generous attention to that wolf is curious; see \textit{Discussion}.

\subsection{\textbf{Referee Report 2}}

The wolf's review of this second pig's manuscript appeared nearly as cursory as the first.  He made no comment regarding the applications of Newton's laws.  Nor did he seem to notice the large fraction of references that cited the work of Little Red Ridinghood's wolf.  His sole note invoked his habitat of Southern California.  He described the alarming escalation of wildfires in this era of climate change, and condemned wooden sticks as a dangerous -- "nay, ignorant and irresponsible" -- choice for building material.  On these grounds, the wolf demanded the major revision of fireproofing, or alternatively, rewriting the entire methodogy using a
\begin{figure*}
  \includegraphics[width=\textwidth]{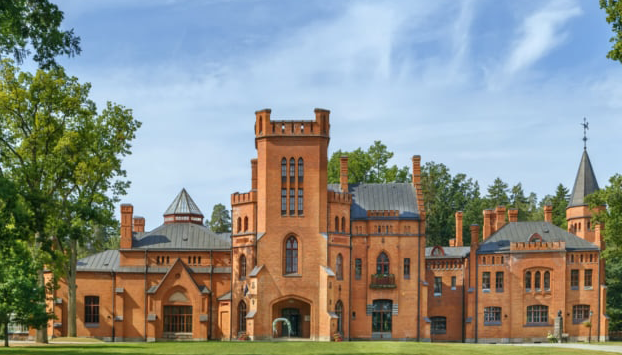}\\
  \caption{\textbf{Third little pig's design for a house of bricks}~\cite{houseBricks}.  Referee's main critique: "Head-in-the-clouds theorist.  Typical."  Recommended unequivocal rejection on the grounds that "He'll never pull that off."}
\label{fig5}
\end{figure*}
\noindent
fire-retardant faux-wood replacement\footnote{This recommendation begs the question: in the first little pig's review, why did the wolf not flag straw as a fire hazard?  See \textit{Discussion.}}.

\section{\textbf{The Third Little Pig}}

\subsection{\textbf{Design and Method}}

The third little pig proposed a house of bricks (Fig.~\ref{fig5}), and established its soundness by conducting a stability analysis to show that the house represented a stable fixed point in its local phase space.  While pigs' excellent cognitive abilities are well documented~\cite{croney2003note,mendl2010pig}, I must confess that this pig's deftness with dynamical-systems-based calculations impressed me.

The pig first defined a deterministic model $\bm{F}$ of the house's dynamics, consisting of dynamical state variables $\bm{bricks}$.  $\bm{F}$ is a map describing a means of forwarding the state of $\bm{bricks}$ in time.  In continuous time $t$, this process is written as:
\begin{align} \label{eq3}
  \bm{F} = \diff{brick_{i}(t)}{t} = f_i(\bm{bricks}),
\end{align}
\noindent which states that the dynamical evolution of each brick $i$ depends on some function $f_i$ acting upon all bricks together.  The operators $f_i$ that act on the bricks are coupled nonlinear relations that represent the external processes of wind, rain, lightning, tsunamis, polar reversals, ornery giants, wicked witches, and dragons:
\begin{align*} \label{eq4}
  f_i \ni (\text{wind, rain, lightning, tsunamis, polar reversals,} \\
  \text{giants, witches, dragons}).
\end{align*}

The little pig then examined the attractors, or states of equilibrium, of this dynamical network of bricks.  The stability of equilibrium states can be characterized via a linear stability analysis~\cite{arnold1973ordinary,glendinning1994stability}: linearizing the functions $f_i$ (of Eq.~\ref{eq3}) about each equilibrium, and constructing a matrix of the partial derivatives of the $f_i$ with respect to the variables $bricks_j$ evaluated at that equilibrium.  This matrix is the Jacobian matrix $\bm{DF}$: 
\begin{align*}
\bm{DF} &= 
\begin{bmatrix}
\large{\frac{\partial f_1}{\partial brick_1}} & \dots & \frac{\partial f_1}{\partial bird_n} \\
\vdots & \ddots & \vdots \\
\frac{\partial f_{n}}{\partial brick_1} & \dots & \frac{\partial f_n}{\partial brick_n},
\end{bmatrix}
\end{align*}
\noindent
where $n$ is the number of bricks.

One way to determine the stability of a particular equilibrium is to examine the spectrum of eigenvalues and eigenvectors of $\bm{DF}$.  The eigenvalues $\lambda$ and corresponding eigenvectors $\bm{v}$ of $\bm{DF}$ are defined by the following relation: 
\begin{align*}
  DF \bm{v}_a &= \lambda_a \bm{v}_a.
\end{align*}  
\noindent
Here, each eigenvector $\bm{v}_a$ represents a direction in the phase space inhabited by the house of bricks.  The eigenvalue $\lambda_a$ is a number associated with eigenvector $\bm{v}_a$ that dictates how trajectories along that direction $\bm{v}_a$ will behave near the equilibrium.  If the phase space is $d$-dimensional, then $a=[1,2,\dots,d]$.  

\begin{figure}
  \includegraphics[width=\textwidth]{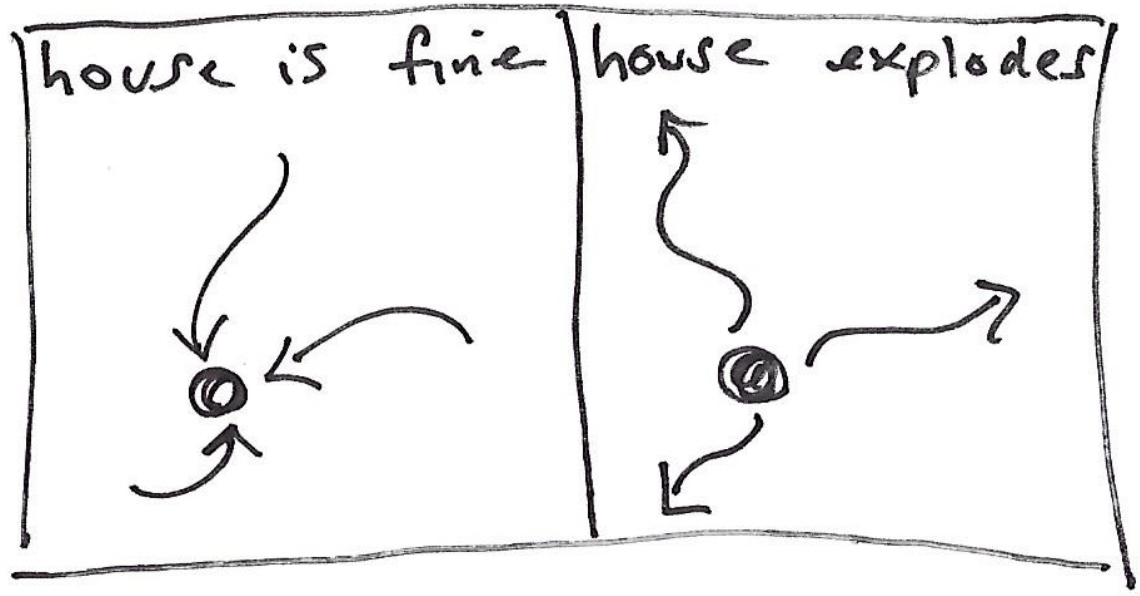}\\
  \caption{\textbf{Third little pig's stability analysis for the house of bricks.}  \textit{Left}: a stable fixed point, where all brick trajectories in the local neighborhood converge to this configuration.  \textit{Right}: an unstable fixed point, where all nearby brick trajectories diverge.  The third little pig claimed, via the analysis described in the text, that the house of bricks could be characterized by the stable fixed point (left).}
\label{fig6}
\end{figure}

If all eigenvalues of $\bm{DF}$ are real and negative, all nearby trajectories will converge to the equilibrium.  That is: the house will stand still (Fig.~\ref{fig6}, left panel).  If all eigenvalues are real and positive, then all trajectories will diverge: the house will explode (Fig.~\ref{fig6}, right panel)~\footnote{Eigenvalues containing imaginary parts were beyond the scope of this third little pig's paper.}.  The pig showed that all eigenvalues were real and negative, and concluded that the proposed house was stable.  

While the pig's analysis was indeed an impressive feat, I am doubtful that he truly used it in order to design the house.  I took it upon myself to research the history of house-, castle-, and fortress-building, and it turns out that many structures built prior to the invention of linear stability analysis -- indeed, even before calculus -- have still managed to weather time quite sportingly~\cite{
brown1955royal,bachrach1983angevin,gardiner2014archaeological,pantin1962medieval,coates1996housing}.  See \textit{Discussion.}  

\subsection{\textbf{Referee Report 3}}

Unfortunately for this third little pig, the big bad wolf turns out to be an expert on dynamical systems and stability analysis.  By contrast to the cursory reports for the first and second little pigs, this third one was a scrutinous and scathing epic.

The wolf's critiques were twofold.  First, he asserted his expert opinion that one does not require a dynamical systems analysis to build a house (thus validating my own suspicion).  Second, given the third little pig's visual depiction (Fig.~\ref{fig5}), he noted that one \textit{does} require hundreds of millions of dollars and an able and willing physical work force, neither of which was addressed in the pig's manuscript.

Regarding the first point, the wolf then delved into specifics.  He faulted the pig for not showing the precise dynamical forms for the coupled nonlinear terms $f_i$ that relate the external processes such as rain, witches, and giants to the brick trajectories.  ("So the reader is asked to just take the author's word for it?  And we are supposed to believe that there exist such things as polar reversals?")

Further, the wolf observed that the giant whom the pig employed in the calculations was "not really that big a giant," that the wicked witch's magical powers were "limited, to say the least," and that the dragon chosen by the pig suffered from severely restrictive asthma.  His point here was that the dynamical model of the house of bricks had not been truly put to the test.  

In addition, and again invoking his Southern California home, he derided the pig for not including earthquakes in the dynamics $f_i$.

The wolf also faulted the third little pig for failing to specify which faraway land he planned to build upon -- an omission that he had not noted in either of the other two pigs' manuscripts~\footnote{In fact, in most fairy tales, the precise location of the faraway land is not specified.}.

The wolf went on to make multiple rather unprofessional comments, including: "What kind of an ego?", "head in the clouds," "full of hot air," and regarding Fig.~\ref{fig5}: "He'll never pull that off."  In summary, the wolf seemed quite determined to doom this third little pig's manuscript.  He recommended unequivocal rejection.

\section{Discussion}

The laudable aim of peer review is that it eliminate bias: that both authors and referees act solely to maintain scientific integrity.  In practice, that is rarely the full story.  With that in mind, let us now examine the peculiar features of both the manuscripts and the reports.

\subsection{\textbf{The pigs: biased by self preservation?}}

Recall that each little pig's manuscript included an oddity.  For the first, second, and third pig, respectively, these were: 1) millet spills and cloven-hoofprints that did not enhance the manuscript; 2) copious citations of papers by Little Red Ridinghood's wolf, who happens to be an expert on home construction; 3) an elaborate yet unnecessary means of analysis.  Now, I do not doubt that the pigs each believed, to varying degrees, in the scientific merit of their work.  To explain these puzzling features, however, I propose the following: that these pigs acted not purely on scientific merit, but also on self preservation.  That is: they heeded their mother's warning that survival in their land requires disseminating news of successful home construction.  This motivation I think somewhat sullied their manuscripts.  

Now, in fact, pigs are well documented to be exploitative and deceptive~\cite{mendl2009learning,held2010domestic,held1t2001studies,mendl2010pig,mendl2009learning}.   In true devious character, then, I propose that they devised conniving fallback plans should their material not pass muster on its own.   Specifically:
\begin{enumerate}
  \item \textit{The first pig added the spilled millet and hoofprints intentionally, with the aim to distract.}  This little pig was well aware of the elementary mathematical content of his manuscript, and worried that it would not meet the journal's standards of rigor.  He hoped to distract the referee from that systemic issue, by dangling minor messes.  
  \item \textit{The second little pig}, knowing that she could not bribe the editor since the journal is not open-access~\footnote{An open-access journal charges authors a fee to publish, which may or may not affect the judgement of the editors who review the submitted manuscripts~\cite{de2018open,kolata2017scholarly}.}, instead \textit{aimed to flatter the referee.}  The copious references she cited were written by Little Red Ridinghood's wolf, a well known expert on building material.  This pig -- as devious as her brother, if not more so -- may have anticipated that her referee would be that particular wolf.  In fact, it is a reasonable guess, given that that wolf would be well qualified to referee her manuscript.  Thus she cited him generously in hopes of currying favor.  This pig's assumption regarding the reviewer's identity might also explain the wooden sticks as her choice of building material: Ridinghood's wolf has demonstrated an affinity for heavily wooded areas.   Unfortunately for her, as the reviewer did not appear swayed by the citations, he was probably not the wolf whom the little pig had anticipated.  In addition, this pig's writing Newton's Second Law in font size six and casting it as "irrelevant" for sticks-based houses strikes me as brazen.
  \item \textit{The third little pig used an ostentatious -- yet unnecessary -- means of analysis in order to dazzle and impress.}  As I admitted, I was so enamored of this pig's calculations that it took me some time to realize that they were not in fact necessary.  I found numerous other well documented brick homes just as impressive as that depicted in  Fig.~\ref{fig5}, constructed without the use of linear stability analysis.  Further, the reviewer -- an expert in stability analysis -- noted himself that its application was unnecessary for this case, commenting that the pig was "full of hot air."  
\end{enumerate}

\subsection{\textbf{The wolf: biased by hunger?}}

Now let us turn to the big bad wolf's motivations.  As noted, some aspects of his reviews merit inspection.  He bestowed upon the straw house an offhand pass, while severely faulting the sticks house for not being fire-proof.  He missed many details contained in the first two papers, while giving the third manuscript a thorough scourge.  To explain these behaviors I propose the following: that at the time he reviewed these manuscripts, the wolf was hungry.

First, hunger is associated with low blood sugar, a physiological state often characterized by difficulty focusing, impatience, and irritability~\cite{gold1995changes}.  This might explain why the wolf missed the first pig's distracting hoofprints and the fact that straw is flammable, his lack of attention to Newton's Second Law, and moreover the cursory nature of the first two reports.  It would also explain why he bristled and lashed out at the third pig, for what the wolf might have considered a conniving exploitation of the dynamical systems framework that lies at the heart of this wolf's own research.

Second, hunger would explain the wolf's relative rankings of the houses' merit.  He accepted the straw house, demanded major revision of the sticks house, and torpedoed the bricks house.  I ask the reader to consider that ranking in light of the wolf's outlook, were one of the little pigs -- one of the fat and tender little pigs, mind you -- to be situated just inside the closed door of one of those houses.

The straw house model, based on toe-counting, appears easily destroyable.   The wolf found no fault with this design.

The model for the sticks house invokes Newton's third law to state that the house would blow back at wind and thereby remain standing.  This statement might have given the wolf pause.  It implies that if the wolf were to huff and puff on the house in order to blow it down, that the house would huff and puff right back, such that the wolf accomplished nothing.  Hence we have his requirement of major revision.

Finally, one glance at Fig.~\ref{fig5} would have made plain to the wolf that all the hot air within his body would not crack that model.  Hence his unequivocal rejection of the manuscript.~\footnote{On a related note, I wonder why the journal did not consider it a conflict of interest to assign a wolf manuscripts written by piglets.  That decision strikes me as negligent.} 

\subsection{\textbf{"Bad" $=$ ?}}

It is worth contemplating why the referee is considered to be the "bad" player here, while the authors are not.  Is the wolf bad because he is hungry and his genetics dictate that he be a carnivore?  Further, are we all not hungry for something, and do we never prioritize that hunger over the interests of others?  

Why do we as authors grow angered at our referees?  Is it because we think they are not paying due attention to our words?  Or because they read our words all too well and thereby force us to attend our faults?

Why do we as referees grow annoyed at the manuscripts we review?  Is it because we think the authors are overstating their feats, or otherwise attempting to sneak something by us?  Is it because they keep confusing the words "that" and "which"?  Or because we feel that our time would be better spent elsewhere? 

Should we be publishing all of these papers?  Do they all merit reading?  As the final chorus of the fairy-tale musical \textit{Into the Woods} tells us: "You can't just act; you have to think"~\cite{intoTheWoods}.  In fact, unfortunately one \textit{can} choose to just act.  In publishing this manuscript, am I serving anyone besides myself?  Do I care?\\

\section{Moral of the Story}

It is unclear whether the pigs or wolf gleaned wisdom from this peer review episode, as they soon moved on to other pursuits.  The moral, then, appears to be that in principle it is a good idea to try to do one's best, although in practice one's best is seldom needed -- and almost never expected.\\ 
\noindent
{\Fontamici{\Huge{A}\Large{nd they all lived tolerably ever after.}}}\\

\hspace{3cm} {\centering{\Fontamici{\Huge{T}\Large{he} \Huge{E}\Large{nd}}}}\\

\vspace{1cm}
\section{Acknowledgements}

This manuscript was inspired by "The Three Little Pigs," whose best known version first appeared in \textit{English Fairy Tales} by Joseph Jacobs in 1890, with Jacobs crediting James Halliwell-Phillipps as the source.  E.~A. acknowledges the National Science Foundation (NSF Grant 2139004), and an Institutional Support for Research and Creativity grant from New York Institute of Technology. 

\nocite{TitlesOn}
\bibliography{refs}

\end{document}